# Spontaneous down conversion in metal-dielectric interface - a possible source of polarization-entangled photons

V. Hizhnyakov, A. Loot

Institute of Physics, University of Tartu, Riia 142, 51014 Tartu, Estonia
E-mail: hizh@fi.tartu.ee

**Abstract.** Spontaneous down conversion in metal-dielectric interface in case of oblique laser excitation is considered. In this case it is possible to generate pairs of polarization entangled surface plasmon polaritons with high efficiency if to use excitation angle corresponding to the Kretschmann configuration for generated plasmons. The efficiency can be additionally enlarged if to use properly nanostructured metallic film allowing one to directly excite surface plasmon polaritons and to fulfill phase matching condition of them with generated surface plasmon polaritons.

## 1. Introduction.

It is expected that quantum communication will constitute an essential part of the interchange of electronic information in future. In order to be in common use this method requires utilization of efficient and miniature sources of polarization entangled photons. Sources of polarization entangled photons at present are mostly based on the spontaneous parametric down conversion (SPDC) - decay of laser photons in crystals without inverse symmetry to photon pairs with the same summary wave vector and frequency [1-8]. The elementary process causing SPDC is the third order quantum process. Schematically it can be presented as

$$1_{\vec{k}_0} \to 1_{\vec{k}_1} + 1_{\vec{k}_2}$$

where $1_{\vec{k}_0}$ denotes an initial (laser) photon with the wave vector $\vec{k}_0$ and the frequency $\omega_0$, $1_{\vec{k}_1}$ and $1_{\vec{k}_2}$ denote the created photons with the wave vectors $\vec{k}_1$ and $\vec{k}_2 = \vec{k}_0 - \vec{k}_1$ (and with the frequencies $\omega_1$ and $\omega_2 = \omega_0 - \omega_1$). One of the generated photons (e.g. $1_{\vec{k}_1}$) is considered as the signal and the other ($1_{\vec{k}_2}$) as the idler. The photons of the signal and the idler may have different polarizations and may be in polarization entangled state which is essential for quantum communication.

In classical limit the down conversion cannot take place if there are no other light except the initial wave. However quantum theory allows the process to take place due to existence of the zero-point fluctuations of the electromagnetic waves. As a result of these fluctuations a photon with the frequency $\omega_0$ can decay into two photons with the same sum frequency. The decay takes place spontaneously. The state of the environment essentially is not changed at SPDC which is the reason to call it as the parametric process. The number of generated pairs of photons depends linearly on the number of initial photons. This means that SPDC has a characteristic yield $\kappa$ - the ratio of created pairs to the number of initial photons which in the low intensity limit depends only on properties of the medium. The value of the yield is usually very low due to extreme weakness of the third-order process with zero-point fluctuations being involved. To observe SPDC, strong laser excitation is required due to what the SPDC-based sources of entangled photons have usually substantial size.

However the probability of SPDC and other nonlinear optical process is possible to increase if to use materials with reduced dimensions. In this way the second-harmonic generation, sum-frequency generation [9-13] and four wave mixing [14] have been studied on different surfaces. Especially strong increase in the efficiency of the nonlinear processes may be achieved if to use the metal-dielectric interfaces. In such a case the surface plasmon polaritons (SPPs) take part in the nonlinear process. These excitations have much smaller group velocity and much stronger electric field than photons in vacuum. The phase matching condition may be here fulfilled by proper excitation or/and nanostructuring of the interface. As a result the efficiency of the nonlinear photons transformations may be increased by many orders of magnitude; see Ref. [14] where a huge enhancement of the four-wave mixing on the nanostructured gold surface has been achieved.

## 2. Yield of spontaneous down conversion in metal-dielectric interface

In this communication we discuss the efficiency of SPDC for surface plasmon polaritons. The process under consideration corresponds to decay of a quantum of surface excitation of the metal-dielectric interface with the frequency $\omega_0$ into two quanta of surface plasmon polaritons with the frequencies $\omega_1$ and $\omega_2 = \omega_0 - \omega_1$. Our consideration is based on the theory of SPDC developed by Klyshko [1, 2]. According to this theory, the radiance of the signal radiation is determined by the second order nonlinear susceptibility of the medium $\chi^{(2)}$ and the pumping intensity $I_0$. In particular, if the thermal fluctuations are negligibly small, the signal radiance (measured in photons per signal-wave mode at the output of the scattering crystal) is given by the equation [2, 7]

$$N_1 = F(k_1)(N_2 + 1) \tag{1}$$

Here $N_2$ is the number of free-field photons in the idler wave mode at the input of the crystal, 1 in the parentheses is the effective radiance of zero-point electromagnetic vacuum fluctuations measured in photons per the idler radiation mode, and $F$ is the parametric transformation coefficient. In a transparent crystal this coefficient is given by the formula

$$F = 4\pi^2 c_0^{-2} \omega_1 \omega_2 |\chi^{(2)}(\omega_1 = \omega_0 - \omega_2)|^2 |E_0|^2 l_\Delta^2 \tag{2}$$

where $\chi^{(2)}(\omega_1 = \omega_0 - \omega_2)$ is the convolution of the second order susceptibility tensor of the crystal with the polarization orts of the pumping, signal, and idler waves; $E_0 = \sqrt{Z_0 I_0}$ V/m is the electric field of the pumping-wave, $Z_0 = 1/c_0 \varepsilon_0 = 376.7$ Ohm is the impedance of free space, $\varepsilon_0$ is the permittivity of free space, $c_0$ is the corresponding speed of light, and $l_\Delta$ is the coherent length. In a homogeneous medium $l_\Delta \approx 1/|\vec{k}_0 - \vec{k}_1^* - \vec{k}_2^*|$ (we take into account that the wave vectors of the contributing waves are complex due to their possible damping). In usual crystals which do not have inverse symmetry $|\chi^{(2)}| \sim 1$ pm/V.

To apply Eq. (2) for the estimation of the efficiency of SPDC in a metal-dielectric interface the second-order susceptibility $\chi^{(2)}$ should be multiplied by the factor $\eta_0\eta_1\eta_2$, where $\eta_0$ stands for the enhancement of the excitation field and $\eta_1$, $\eta_2$ stands for the enhancement of the fields of emitted quanta. SPPs in metal-dielectric interfaces have much smaller group velocity than photons in vacuum and in usual crystals allowing to enhance the local field and in this way the nonlinear interaction.

Below we consider creation of pairs of quanta of SPPs with the frequencies $\omega_1 = \omega_2 = \omega_0/2$. Taking into account that the excitation power equals $I_0 = N_0\hbar\omega_0^2$, the yield of corresponding SPDC can be presented in the form:

$$\kappa \equiv \frac{N_1}{N_0} = \alpha\eta_0^2\eta_1^4\hbar\omega_0^2 Z_0 \frac{l_\Delta^2}{\lambda^4}\left|\chi^{(2)}\right|^2 \qquad (3)$$

where $N_0$ is the number of photons of excitation, $\alpha \sim 10^{-2}$ is the dimensionless parameter, $\lambda = 4\pi c_0/\omega_0$ is the wave length of the generated photons (Eq. (3) is presented in the form allowing one to easily see that $\kappa$ is a dimensionless quantity: the dimensionality of $\hbar\omega_0^2$ is W, the dimensionality of $\hbar\omega_0^2 Z_0/\lambda^2$ is (V/m)$^2$, i.e. the opposite to the dimensionality of $\left|\chi^{(2)}\right|^2$). The yield is the same for continuous and pulse lasers excitation supposing that the propagation direction is the same and the coherence length of the laser light is larger than $l_\Delta$.

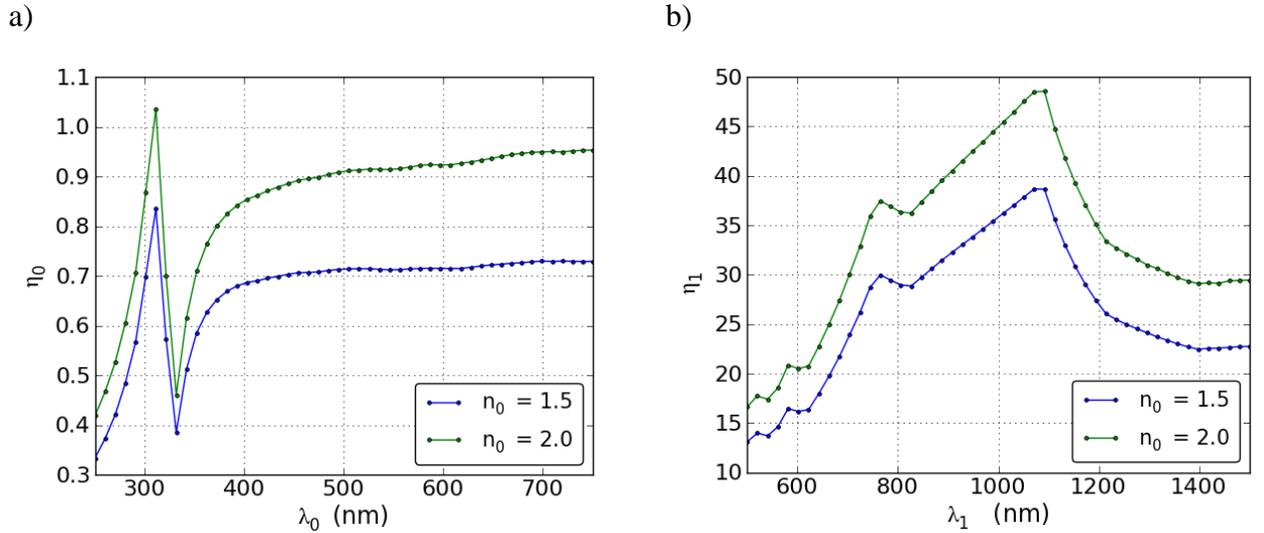

Fig. 1. The renormalization (enhancement) of excitation (a) and emitted (b) fields for SPDC in a structure consisting of prism (refractive index n$_0$), silver film with thickness 60 nm and vacuum. The excitation angle for both figures (a), (b) corresponds to Kretschmann configuration for surface plasmon resonance with frequency $\omega_1 = 2\pi c_0/\lambda_1$.

In case of SPDC in usual crystals, where no remarkable enhancement of the fields take place ($\eta_0\eta_1\eta_2 = 1$) for $\left|\chi^{(2)}\right| \sim 1$ pm/V - an usual value of the second-order susceptibility in crystals without inverse symmetry, $\lambda = 1$ μm and $\lambda_\Delta = 1$ mm one gets $\kappa \sim 10^{-12}$, which is the well-known estimation of the yield of ordinary SPDC (see, e.g. [15]). However in metal-dielectric interface it is possible to generate pairs of SPP quanta. This allows one to achieve strong enhancement of the local field and in this way to strongly increase the nonlinear interaction under consideration. In Kretschmann prism-metal-air configuration, it is easy to calculate field enhancements $\eta_0$ and $\eta_1$ from 3-layer Fresnel transmission coefficient or by transfer-matrix method [18]. The results for silver are presented on the Fig. 1, where both $\eta_0$ and $\eta_1$ are calculated. The excitation angle for both figures Fig. 1a, 1b are the same and corresponds to the surface plasmon resonance angle for frequency $\omega_1 = 2\pi c_0/\lambda_1$. In other words, it corresponds to Kretschmann configuration for surface plasmon resonance angle in figure (b). The calculations are done for two different prisms with refractive index $n_0=1.5$ and $n_0=2.0$. The refractive index of silver film is taken form Ref. [17] and the thickness is 60 nm (optimal). The enhancement factor $\eta_1$ for the SPPs in the near-infrared region is close to 45 (see Fig. 1). The coherence length in this case may reach the value $l_\Delta \approx 1/2 \operatorname{Im} k_{\omega_1}$. In the red – near infrared region $\operatorname{Im}\left[ k_{\omega_1} \Big/ \dfrac{2\pi}{\lambda_0} \right] = \operatorname{Im}\left[ n_{sp} \right] \approx 10^{-4}$ (see Fig. 2), which means that $l_\Delta$ may reach few mm. This means that using SPPs one can get $\kappa > 10^{-5}$.

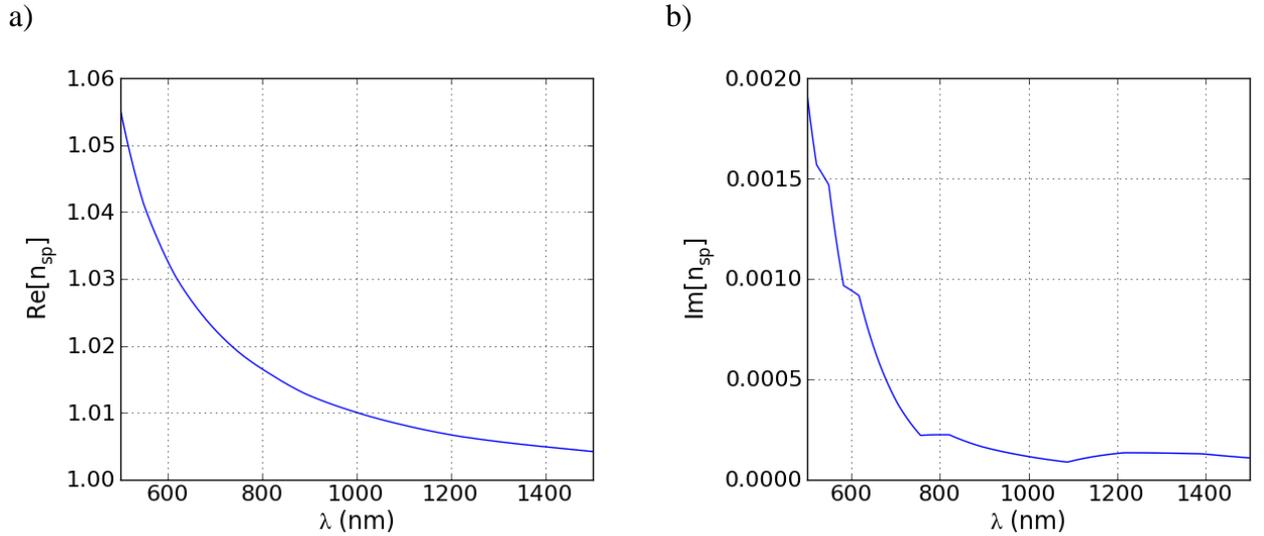

Fig.2. Real (a) and imaginary (b) parts of effective mode index $n_{sp}$ of surface plasmon polaritons at the air-silver (refractive index form [17]) interface as a function of the wavelength of photons.

## 3. Scheme of setup

A scheme of setup is presented on Fig. 3. According to this scheme a silver film is placed between Kretschmann prism and a dielectric plate. The film is excited by the p-polarized light with the frequency $\omega_0 = 2\omega_1$ through the prism with the angle $\varphi_0 = \arccos(n_{sp}/n_0)$, where $n_0$ is the refractive index of prism and $n_{sp} = Re(k_{\omega 1})\lambda_0/2\pi$ is the effective refractive index of the surface plasmon polaritons in the interface with the frequency $\omega_1$. This angle corresponds to the Kretschmann configuration for the light with the frequency $\omega_1 = \omega_0/2$.

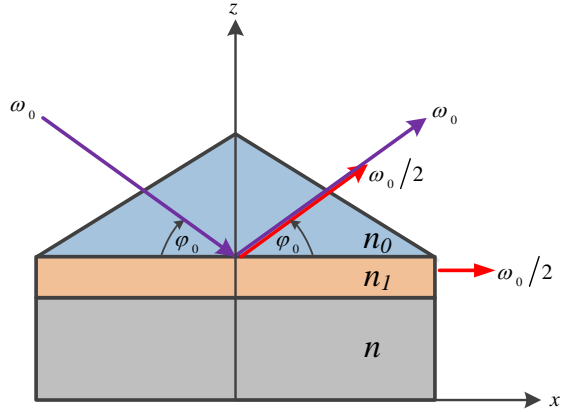

Fig. 3. Spontaneous parametric down conversion in the interface of dielectric prism, metallic film and dielectric plate. The angle $\varphi_0$ of laser excitation corresponds to the Kretschmann configuration for the frequency $\omega_1 = \omega_0/2$ of generated surface plasmon polaritons. The emission of photon pairs with the frequency $\omega_0/2$ takes place a) from the prism at the same angle as the reflected light and b) forward from the end of the interface.

The excitation creates the polarization in the interface, which propagates along the interface with the phase velocity $\omega_0 n_0 \cos\varphi_0 = c_0/n_{sp}$ being equal to the phase velocity of plasmons with the frequency $\omega_1 = \omega_0/2$ (corresponding wave vector is $\vec{k}_0 n_0 \cos\varphi_0$). In the scheme under consideration $\chi^{(2)}$ differs from zero due to lack of symmetry with respect to reflection from the surface of the interface ($z$ direction in Fig. 3). Therefore the running wave of polarization with the frequency $\omega_0$ can generate the pairs of plasmons with the same sum frequency. The phase matching condition

$$\vec{k}_0 n_0 \cos\varphi_0 = \vec{k}_1 + \vec{k}_2 \qquad (4)$$

is fulfilled for both plasmons having the same wave vector $\vec{k}_1 = \vec{k}_2 = \vec{k}_0 n_0 \cos\varphi_0/2$ (and frequency $\omega_1 = \omega_2 = \omega_0/2$). The generated plasmons can transfer to photons in the Kretschmann prism with the propagating angle $\pi - \varphi_0$ and also at the end of the interface (see Fig. 3).

To estimate the possible yield of SPDC in this scheme one should take into account that the laser light, due to the phase mismatch does not create the SPPs with the frequency $\omega_0$. However it creates the running wave along the interface of forced oscillations of the electromagnetic field. Obviously this field is not enhanced (Fig. 1a). However the fields of created quanta of SPPs are enhanced (Fig. 1b). Therefore the enhancement factor in the case under consideration is of the order of $\eta_1^4$. In case of silver film of the thickness 60 nm and Kretschmann prism with the refractive index 1.5 the enhancement factor for SPPs with the frequency $\omega_1 = 1.88$ fs$^{-1}$ equals $\eta_1 = 35$; see Fig. 1 (corresponding wave has 1 µm wavelength in vacuum). The imaginary part of the wave number of these SPPs is one - two cm$^{-1}$ (see Fig. 2) allowing one to reach the coherent length $l_\Delta \approx 3$ mm. Taking once more $\chi^{(2)} \sim 1$ pm/V we find that in the order of magnitude $\kappa \sim 10^{-4}$. This is very large yield as compared to referred above yield of SPDC in ordinary crystals without inverse symmetry.

## 4. Polarization entanglement of emitted photon pairs

Presented experimental scheme allows one to generate polarization entangled photons. Indeed in this scheme the mirror symmetry in $z$ direction is absent. Therefore the components $\chi^{(2)}_{y,yz}$ and $\chi^{(2)}_{y,zy}$ of the second-order nonlinear susceptibility tensor describing two-photon emission in $x$ direction differ from zero (here the first index corresponds to polarization of excitation, while last two indexes correspond to the polarization of emitted quanta). This means that one of photons in the emitted pair of photons has $y$ polarization and another has $z$ polarization. Due to identity of both emitted photons their wave function must be the linear combination of the wave functions with both polarizations

$$|\psi\rangle_{yz} = \frac{1}{2}\left(|\psi_{1z}\rangle|\psi_{2y}\rangle + |\psi_{2z}\rangle|\psi_{1y}\rangle\right) \qquad (5)$$

where $|\psi_{n\alpha_0}\rangle$ is the wave function of a photon number $n=1,2$ with polarization index $\alpha_0$. This wave function of the pair of emitted photons describes the basic for the quantum informatics Bell-type state [1]. This state cannot be presented as a product of two one-photon states (otherwise there will be a possibility to find both photons with the same polarization which is impossible here). This means that the emitted photons are in the polarization entangled state. If to divide photons directionally and to choose the polarization of one of them (signal) to be, e.g. $z$, then polarization of another photon (idler) will be $y$. However if to choose polarization of the signal photon to be $y$ then the idler photon will be polarized in $z$ direction. It is just what one needs for using polarization entangled photons for quantum cryptography. For better distinguishing of signal and idler the excitation with angle $\varphi < \varphi_0$ may be used. In this case, as it is noticed above the photon pairs consists of photons with different wave vectors.

## 5. Discussion

According to our consideration the proposed setup here allows one to prepare a source of polarization entangled photons with the good yield $\kappa \sim 10^{-4}$. One can achieve even larger yield if to use properly nanostructured metal film (see Fig. 4), in the same way as it was done in Ref. [4] for four wave mixing. Indeed, a grating of the metallic film of the interface with the period $a$ results in appearance of SPPs with the same frequency but with the new wave numbers $k_n = k - nk_a$, $n = \pm 1, \pm 2,...$ For properly chosen $a$ the laser wave with the longitudinal component $\omega_0 \cos\varphi_0 / c_0$ of its wave number can excite a SPPs with the frequency $\omega_0$ in the interface (e.g. the one with the wave number $\omega_0 \cos\varphi_0 / c_0 + nk_a$). This will result in strong enhancement of the moving polarization allowing one to get $\eta_0 \gg 1$. This make possible to increase additionally the yield of the SPDC more than two orders of magnitude.

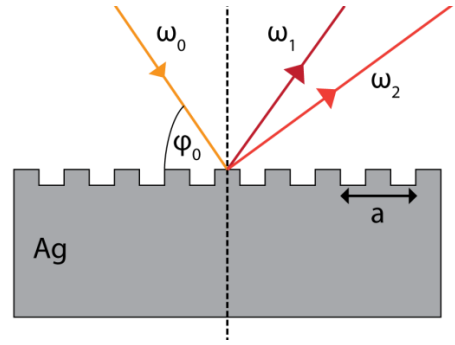

Fig. 4 Spontaneous parametric down conversion at the interface of dielectric and nanostructured metallic film.

The presented scheme may be also modified so that the photons in the generated pairs will propagate slightly differently and/or will have slightly different frequencies. This will be achieved if to use excitation with the angle $\varphi$ slightly less than $\varphi_0$. In this case the phase matching condition

$$\vec{k}_0 n_0 \cos\varphi = \vec{k}_1 + \vec{k}_2 \tag{6}$$

will be fulfilled for $\vec{k}_1 \neq \vec{k}_2$, where $|\vec{k}_0| = \omega_0/c_0$ is the wave number of the laser light.

If to use excitation with the angle $\varphi > \varphi_0$ then the SPDC will not take place due to mismatch of phases of excited by laser polarization and plasmons with the same sum frequency: in this case the excited polarization moves along the interface faster (superluminally) than plasmons. However in this case two-photon emission is also possible, however due to another mechanism – the dynamical Casimir effect. Generation of pairs of entangled photons in metal-dielectric interface in case of excitation with $\varphi > \varphi_0$ was studied in [16].

Finally we note that presented above estimation of the yield of SPDC is made for a perfectly made setup. In real experiment the losses are expected which will result in reduction of the yield one or two orders of magnitude. Therefore the presented estimation of $\kappa$ is accurate only up to one – two orders of magnitude. However it does not alter our conclusion that the value of $\kappa$ for the plasmon-supported SPDC should be much higher than in case of usual SPDC experiments.

### Acknowledgements

The research was supported by Estonian research projects SF0180013s07, IUT2-27 and by the European Union through the European Regional Development Fund (project 3.2.0101.11-0029).

### References

[1] D. N. Klyshko, Pis'ma Zh. Éksp. Teor. Fiz. 6, 490 (1967) [JETP Lett. 6, 23 (1967)].
[2] David. N. Kyshko, *Photons and Nonlinear Optics*, Gordon and Breach Science Publishers, New York, London, Paris, Monreux,Tokyo, Melbourne, 1988.
[3]. S. A. Akhmanov, V. V. Fadeev, R. V. Khokhlov, and O. N. Chunaev, Pis'ma Zh. Éksp. Teor. Fiz. 6, 575 (1967) [JETP Lett. 6, 85 (1967)].
[4]. S. E. Harris, M. K. Oshman, and R. L. Byer, Phys. Rev. Lett. 18, 732 (1967)
[6] C. Kurtsiefer, M. Oberparleiter and H. Weinfurter, *Generation of correlated photon pairs in type-II parametric down conversion – revisited,* Journal of Modern Optics, vol. 48, pp. 1997-2007, 2001.

[7] G. Kh. Kitaeva and A. N. Penin, *Spontaneous Parametric Down-Conversion*, JETP Letters, Vol. 82, No. 6, 2005.
[8] R.W. Boyd, Nonlinear Optics, Academic Press, San Diego, 2008, 3rd ed.
[9] F. Brown, R. E. Parks, and A. M. Sleeper, Phys. Rev. Lett.14, 1029 (1965).
[10] N. A. Papadogiannis, P. A. Loukakos, and S. D.Moustaizis, Opt. Commun. 166, 133 (1999).
[11] B. Lamprecht, J. R. Krenn, A. Leitner, and F. R.Aussenegg, Phys. Rev. Lett. 83, 4421 (1999).
[12] S. Palomba and L. Novotny, Phys. Rev. Lett. 101, 056802(2008).
[13] A. Leitner, Mol. Phys. 70, 197 (1990).
[14] A. Bouhelier, M. Beversluis, A. Hartschuh, and L.Novotny, Phys. Rev. Lett. 90, 013903 (2003).
[15] Alexander Ling, Antía Lamas-Linares, and Christian Kurtsiefer, *Absolute emission rates of spontaneous parametric down-conversion into single transverse Gaussian modes*, PRA 77, 043834 (2008).
[16] V. Hizhnyakov, *Plasmon-supported emission of entangled photons and zero-point energy*, arXiv:1311.6824 [physics.optics].
[17] P.B Johnson and R. W. Christy, *Optical constants of the noble metals*, Phys. Rev. B 6, 4370–4379 (1972).
[18] L. Novotny and B. Hecht, *Principles of Nano-Optics*, Cambridge University Press, 2006.